\documentclass[12pt,a4paper]{article}
\usepackage{lineno,hyperref}
\usepackage[latin5]{inputenc}
\usepackage{amsmath}
\usepackage{graphicx}
\usepackage{subfigure}
\usepackage{float}
\usepackage{mathtools}
\usepackage{color}
\usepackage{cite}
\usepackage{mathrsfs}
\usepackage[]{geometry}
\usepackage[singlelinecheck=false]{caption} % tablo basl�g� sola dayal�
\usepackage{graphicx}
\usepackage{cite}
\usepackage{authblk}
\usepackage{float}
\usepackage{color}
\usepackage{graphicx}
\usepackage{epstopdf}
\usepackage{amscd,amsmath,amssymb,amsthm,amsfonts,epsfig,graphics}

\newtheorem{theorem}{Theorem}
\title{Topological 1-soliton solutions to some conformable fractional partial differential equations}
\author{G\"{o}khan Koyunlu \thanks{g.koyunlu@nileuniversity.edu.ng}\\ 
Department of Computer Engineering\\
 Nile University of  Nigeria. Abuja, Nigeria.}

\begin{document}
\maketitle
\begin{abstract} % abstract
Topological 1-soliton solutions to various conformable fractional PDEs in both one and more dimensions are constructed by using simple hyperbolic function ansatz. Suitable traveling wave transformation reduces the fractional partial differential equations to ordinary ones. The next step of the procedure is to determine the power of the ansatz by substituting  it into the ordinary differential equation. Once the power is determined, if possible, the power determined form of the ansatz is substituted into the ordinary differential equation. Rearranging the resultant equation with respect to the powers of the ansatz and assuming the coefficients are zero leads to an algebraic system of equations. The solution of this system gives the relation between the parameters used in the ansatz.
\end{abstract}
%\begin{keyword}
%PDE, symbolic pro, solitary waves
\textbf{\textit{AMS2010:}} 5C07;35R11;35Q53.\\%Traveling wave solutions-fractionalPDEs-KdV-likeeqns}
%\end{keyword}
%\end{frontmatter}

%\linenumbers
\section{Introduction}
In the literature of fractional partial differential equations, several definitions are used to construct the equations. One of the recent definitions is the conformable fractional derivative. The main description and some useful properties of this derivative is given in the next section.

This new definition of the fractional derivative has been used to derive new forms, probably a more general ones, of the nonlinear partial differential equations. Thus, exact solutions of these equations have become more significant in both theoretical studies and real world applications. There exist plenty of techniques in the literature to solve nonlinear partial differential equations. Lately, most of them have been adapted to solve fractional partial differential equations when possible.

When the exact solutions to nonlinear partial differential equations are examined, one recognize that a method may be sufficient to generate an exact solutions to some equations but does not guarantee exact solutions to all nonlinear partial differential equations. The non linearity is a general concept and characteristics of each non linearity can be completely different from the other. This perspective of non linearity forces researchers to implement the known methodsof solution to each non linear partial differential equations. 

In the literature, there are many classical methods to solve non linear partial differential equations to set up exact solutions of various types. Simple equation methods \cite{zayed1,roshid1,kaplan1,kaplan2}, first integral method \cite{eslami1,lu1,bekir1}, variations of Kudryashov methods \cite{demiray1,aksoy1,korkmaz1}, ($G'/G$-)expansion methods \cite{guner1,ganji1,naher1},and  various hyperbolic function ansatz techniques \cite{guner2,korkmaz2,guner3} can be listed as some of well known methods to develop exact solutions to non linear partial differential equations. These techniques can also be implemented to fractional partial differential equations for some types of  equations. 

This study aims to implement hyperbolic tangent ansatz methods to develop topological 1-soliton solutions to some conformable time fractional partial differential equations in one and two space dimensions, namely the fractional modified Equal-Width (fmEW), the fractional Klein-Gordon (fKG), and the fractional potential  Kadomtsev-Petviashvili (fpKP) equations. All the fractional derivatives used in these equations are chosen as newly defined conformable derivative. In order not to make restrictions in space domain, the only time fractional forms of the equations are considered throughout the study.

Before setting up the topological 1-soliton solutions, some significant properties of conformable fractional derivative are briefly described in the next section. The considered equations are described briefly and topological 1-soliton solutions are developed in the following sections.

\section{Conformable Derivative}
The conformable fractional derivative definition is given by Khalil \cite{khalil1} as
\begin{equation}
D^{\alpha}(u(t))=\lim\limits_{h\rightarrow 0}{\dfrac{u(t+h t^{1-\alpha})-u(t)}{h}}
\end{equation}
where $\alpha \in (0,1]$ and $u:[0,\infty)\rightarrow \mathbb{R}$ in the half space $t>0$. This fractional derivative supports plenty of properties given below under the assumptions that the order is $\alpha \in (0,1]$ and that $u=u(t)$ and $v=v(t)$ are sufficiently $\alpha$-differentiable for all $t>0$. Then,

\begin{itemize}
\item $D^{\alpha}(a_1u+a_2v)=a_1D^{\alpha}(u)+a_2D^{\alpha}(v)$ 
\item $D^{\alpha}(t^k)=kt^{k-\alpha}, \forall k \in \mathbb{R}$
\item $D^{\alpha}(\lambda)=0$, for all constant $ \lambda$
\item $D^{\alpha}(uv)=uD^{\alpha}(v)+vD^{\alpha}(u)$
\item $D^{\alpha}(\frac{u}{v})=\dfrac{vD^{\alpha}(u)-uD^{\alpha}(v)}{v^2}$
\item $D^{\alpha}(u)(t)=t^{1-\alpha}\frac{du}{dt}$
\end{itemize}
for $\forall a_1,a_2 \in \mathbb{R}$\cite{atangana1,cenesiz1}. The conformable derivative gives support to Laplace transformations, exponential function properties, chain rule, Taylor Series expansion, etc. \cite{abdeljawad1}. Probably the most useful property indicates the relation between the conformable derivative and common derivative. 
\begin{theorem}
Let $u$ be an $\alpha$-conformable differentiable function, and $v$ is also differentiable function defined in the range of $u$. Then,
\begin{equation}
D^{\alpha}(u\circ v)(t)=t^{1-\alpha}v^{\prime}(t)u^{\prime}(v(t))
\end{equation}
\end{theorem}

\section{The Main Frame of Method}
Consider a fractional order partial differential equation in a general form
\begin{equation}
F_1(u,D_{t}^{\alpha}u,u_x,D_{t}^{\alpha}u_x,u_{xx},\ldots )=0 \label{fpde}
\end{equation}
The traveling wave transformation
\begin{equation}
	u(x,y,t)=U(\xi), \xi = ax+by-\frac{\nu}{\alpha}t^{\alpha} \label{wt}
\end{equation}
reduces the fractional partial differential equation (\ref{fpde}) to
\begin{equation}
	F_2(U,U',U'',\ldots )=0 \label{oode}
\end{equation}
where $ ' $ denotes the ordinary derivative. In the traveling wave transformation (\ref{wt}), $a$, $b$, and $\nu$ are assumed as constants. The next step of the method is to suppose that (\ref{oode}) has a solution of the form
\begin{equation}
	U(\xi)=A\tanh^B{\xi} \label{solode}
\end{equation}
where $A\neq 0$ and $B\in \mathbb{Z}^{+}$ are constants to be determined. Substituting this solution into (\ref{oode}) and rearranging the resultant algebraic form for the powers of $\tanh$ function gives an equation. The power constant $B$ is determined by equating the powers of the $\tanh$ function, if possible. After determination of $B$, the solution (\ref{solode}) is substituted into (\ref{oode}) with writing the value of $B$. Rearranging the resultant equation for powers of $\tanh$ and equating the coefficients of them to zero give an algebraic equation system to be solved for $A$, $\nu$, $a$, $b$. 

Note that when the dimension of space domain of $u$ is $2$, then, one of $a$, $b$  is removed from the wave transformation (\ref{wt}). Similarly, the wave transformation (\ref{wt}) is modified in a compatible form while studying in one space dimension.

\section{Topological 1-Soliton Solutions to Time Fractional Modified EW (fmEW) Equation }
The fmEW equation is given as
\begin{equation}
D_{t}^{\alpha}u(x,t)+pu(x,t)u_x(x,t)+qD_{t}^{\alpha}u_{xx}(x,t)=0 \label{ew}
\end{equation}
where $D_{t}^{\alpha}$ is the $\alpha$th order conformable fractional derivative and the subscripts denote the ordinary derivative. The traveling wave transformation (\ref{wt}) for $b=0$ converts (\ref{ew}) to
\begin{equation}
	-\nu U'+paU^2U'-q\nu a^2U'''=0
\end{equation}
Integrating this equation once leads
\begin{equation}
	-\nu U+\frac{1}{3}paU^3-q\nu a^2U''=K \label{odee}
\end{equation}
where $K$ is constant of integration. Substituting the hyperbolic tangent ansatz $\tanh ^B (\xi)$ into (\ref{odee}) gives
\begin{equation}
\begin{aligned}
	&\frac{1}{3}{A}^{3}ap \tanh ^{3\,B} \xi +  \left( 2\,\nu\,q{a}^{2}A{B}^{2}-\nu\,A \right)  \tanh ^{B} \xi
	+ \left( -\nu\,q{a}^{2}A{B}^{2}+AB{a}^{2}\nu
\,q \right)  \tanh^{B-2} \xi \\
 &+ \left( -
\nu\,q{a}^{2}A{B}^{2}-AB{a}^{2}\nu\,q \right)  \tanh^{B+2}  \xi=K
\end{aligned}
\end{equation}
Equating the powers $3B=B+2$ results $B=1$. Thus, substituting $\tanh\xi$ into (\ref{odee}) results
\begin{equation}
	 \left( 1/3\,pa{A}^{3}-2\,{a}^{2}\nu\,Aq \right) \tanh^{3}  
\xi  + \left( 2\,{a}^{2}\nu\,Aq-\nu\,A \right) 
\tanh  \xi  -K=0
\end{equation}
Since $\tanh\xi$ is a nonzero solution,then, the coefficients of $\tanh \xi$ and $\tanh^3\xi$ should be zero in addition to the zero integration constant, $K=0$. Thus, the solution of the system of algebraic equations
\begin{equation}
	\begin{aligned}
	1/3\,pa{A}^{3}-2\,{a}^{2}\nu\,Aq&=0 \\
	2\,{a}^{2}\nu\,Aq-\nu\,A&=0
	\end{aligned}
\end{equation}
gives
\begin{equation}
		\begin{aligned}
	A&=\pm \sqrt {3{\frac {\sqrt {2}\sqrt {{q}^{-1}}\nu\,q}{p}}} \\
	a&= 1/2\,\sqrt {2}\sqrt {{q}^{-1}}
	\end{aligned}
\end{equation}
and
\begin{equation}
		\begin{aligned}
	A&=\pm \sqrt {\pm 3{\frac {\sqrt {2}\sqrt {{q}^{-1}}\nu\,q}{p}}} \\
	a&=- 1/2\,\sqrt {2}\sqrt {{q}^{-1}}
	\end{aligned}
\end{equation}
for arbitrarily chosen $\nu$. These values of $A$ and $a$ gives several solutions to (\ref{odee}) as
\begin{equation}
\begin{aligned}
	U_{1,2}(\xi)&=\pm \sqrt { 3{\frac {\sqrt {2}\sqrt {{q}^{-1}}\nu\,q}{p}}} \tanh{\xi} \\
	U_{3,4}(\xi)&=\pm \sqrt {- 3{\frac {\sqrt {2}\sqrt {{q}^{-1}}\nu\,q}{p}}} \tanh{\xi}
	\end{aligned}
\end{equation}
Returning the original variables makes the solutions
\begin{equation}
\begin{aligned}
	u_{1,2}(x,t)&=\pm \sqrt {3{\frac {\sqrt {2}\sqrt {{q}^{-1}}\nu\,q}{p}}} \tanh{\left( 1/2\,\sqrt {2}\sqrt {{q}^{-1}} x- \nu \frac{t^{\alpha}}{\alpha}\right)}\\
	u_{3,4}(x,t)&=\pm \sqrt {- 3{\frac {\sqrt {2}\sqrt {{q}^{-1}}\nu\,q}{p}}} \tanh{\left( - 1/2\,\sqrt {2}\sqrt {{q}^{-1}} x- \nu \frac{t^{\alpha}}{\alpha}\right)}
		\end{aligned}
\end{equation}
for $p\neq 0$ and $q \neq 0$.

\begin{figure}[H]
 
   \subfigure[$\alpha=0.25$]{
   \includegraphics[scale =0.4] {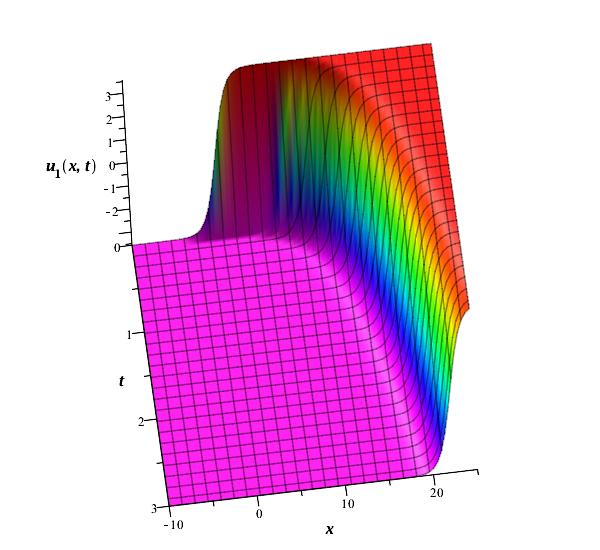}
   \label{fig1a}
 }
  \subfigure[$\alpha=0.50$]{
   \includegraphics[scale =0.4] {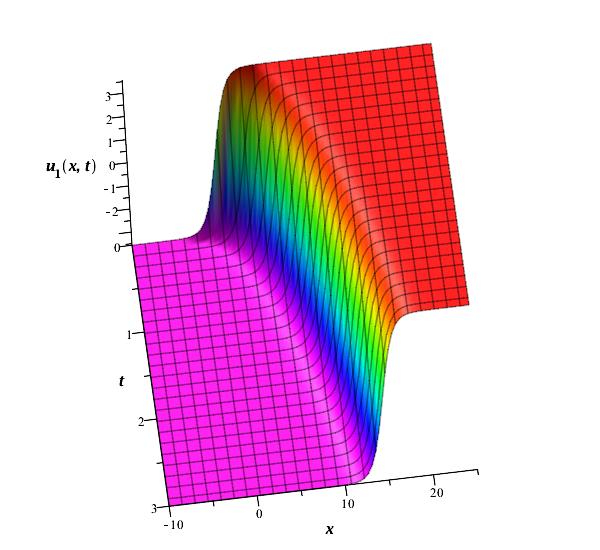}
   \label{fig1b}
 }
 \subfigure[$\alpha=0.75$]{
   \includegraphics[scale =0.4] {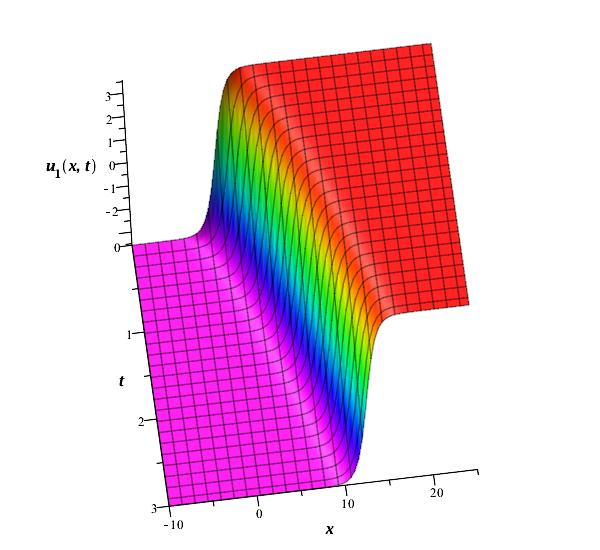}
   \label{fig1c}
 }
 \subfigure[$\alpha=1.00$]{
   \includegraphics[scale =0.4] {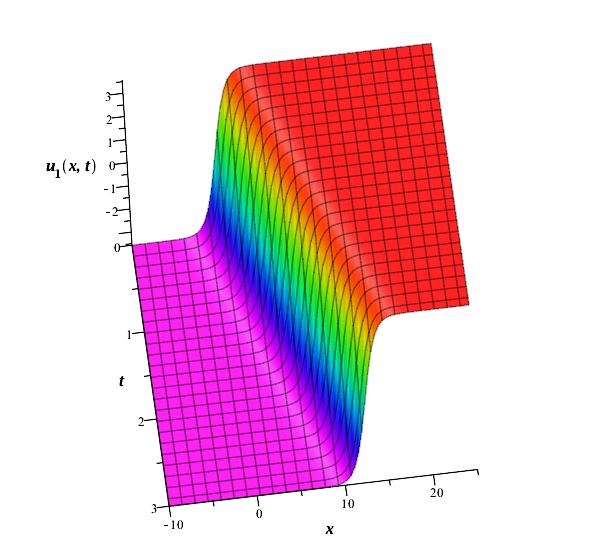}
   \label{fig1d}
 }
 \caption{Illustration of the solutions $u_1(x,t)$ for $p=q=1$ and $\nu =3$}
\end{figure}

The graphical illustration of $u_1(x,t)$ for particular choices of $p$, $q$ and $\nu$ is given in Fig \ref{fig1a}-\ref{fig1d} for various values of $\alpha$.

\section{Topological 1-Soliton Solutions to Time Fractional Klein-Gordon Equation }
\begin{figure}[H]
 
   \subfigure[$\alpha=0.25$]{
   \includegraphics[scale =0.4] {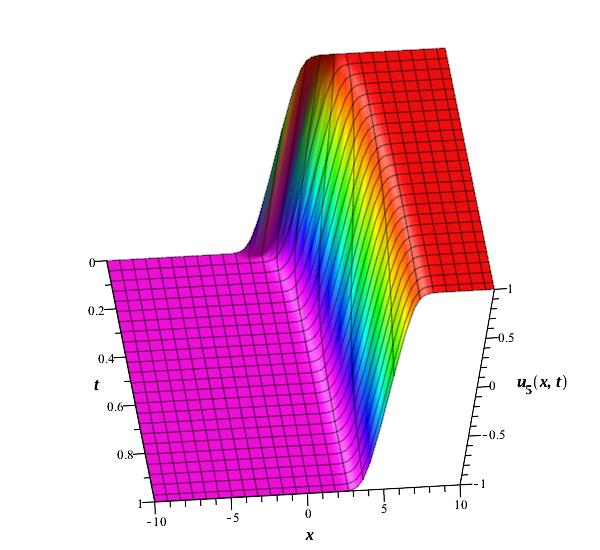}
   \label{fig2a}
 }
  \subfigure[$\alpha=0.50$]{
   \includegraphics[scale =0.4] {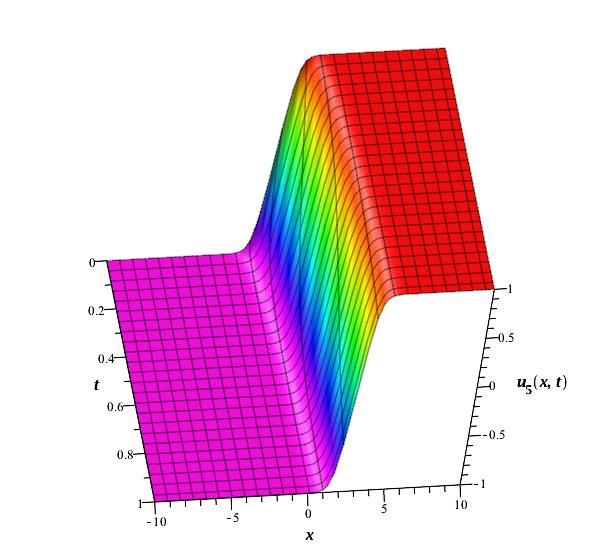}
   \label{fig2b}
 }
 \subfigure[$\alpha=0.75$]{
   \includegraphics[scale =0.4] {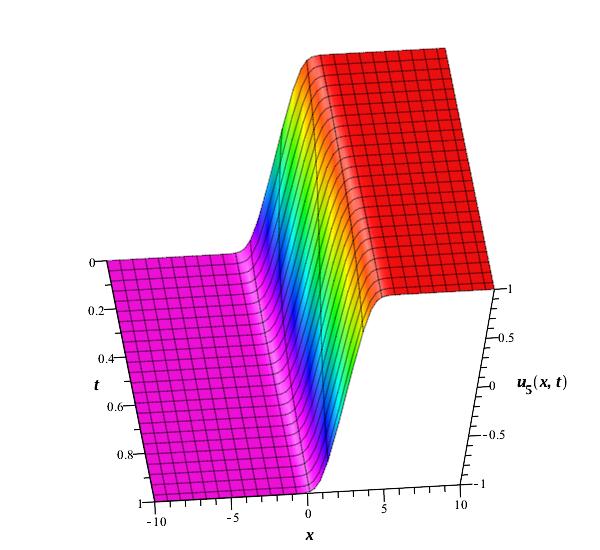}
   \label{fig2c}
 }
 \subfigure[$\alpha=1.00$]{
   \includegraphics[scale =0.4] {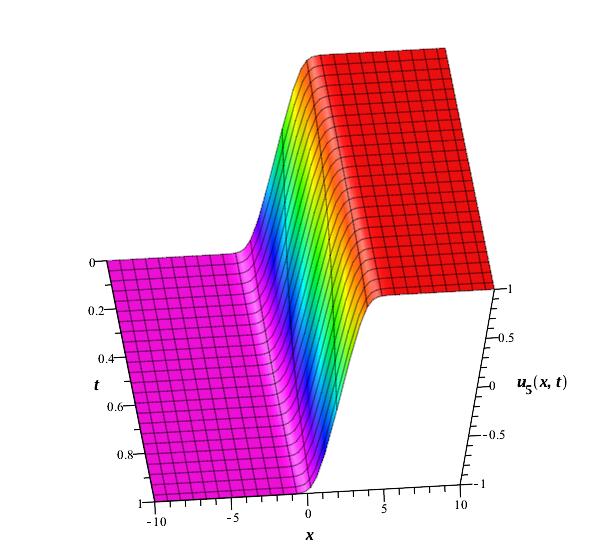}
   \label{fig2d}
 }
 \caption{Illustration of the solutions $u_5(x,t)$ for $p=-1$, $q=1$ and $\nu =2$}
\end{figure}

The conformable fractional Klein-Gordon equation
\begin{equation}
	D_{t}^{2\alpha}u-u_{xx}-pu-qu^3=0
\end{equation}
can transformed into 
\begin{equation}
	\nu ^2U''-a^2U''-pU-qU^3=0 \label{kg}
\end{equation}
via the compatible wave transformation (\ref{wt}) for $\xi =ax-\nu \frac{t^{\alpha}}{\alpha}$. Substituting $\tanh^{B}\xi$ into \ref{kg} gives
\begin{equation}
	\begin{aligned}
	&-{A}^{3}q \tanh^{3\,B} \xi  +\left( -A{B}^{2}{a}^{2}+A{B}^{2}{\nu}^{2}+A
B{a}^{2}-AB{\nu}^{2} \right)   \tanh^{B-2}  \xi   \\
&+\left( 2\,
A{B}^{2}{a}^{2}-2\,A{B}^{2}{\nu}^{2}-pA \right)  \tanh^{B}  
\xi   
 + \left( -A{B}^{2}{a}^{2}+A{B}^{2}{\nu}^{2}-AB{a}^{2}+A
B{\nu}^{2} \right) \tanh ^{B+2} \xi =0
	\end{aligned}
\end{equation}
Equating $3B=B+2$ gives $B=1$. When the solution $\tanh \xi$ is substituted into (\ref{kg}), it takes the form
\begin{equation}
	 \left( -{A}^{3}q-2\,A{a}^{2}+2\,A{\nu}^{2} \right)   \tanh^{3}
 \xi + \left( 2\,A{a}^{2}-2\,A{\nu}^{2}-pA
 \right) \tanh  \xi  =0
\end{equation}
When the coefficients are equated to zero, the algebraic system
\begin{equation}
	\begin{aligned}
	 -{A}^{3}q-2\,A{a}^{2}+2\,A{\nu}^{2} &=0\\
	2\,A{a}^{2}-2\,A{\nu}^{2}-pA&=0
	\end{aligned}
\end{equation}
Solving this system for $A$ and $a$
\begin{equation}
	\begin{aligned}
	A&=\pm\sqrt{-\frac{p}{q}} \\
	a&=\pm\sqrt{\nu^2+\frac{1}{2}p}
	\end{aligned}
\end{equation}
Thus, the solutions to (\ref{kg}) is determined as
\begin{equation}
	\begin{aligned}
	U_{5,6}(\xi)= \pm\sqrt{-\frac{p}{q}} \tanh(\xi)
	\end{aligned}
\end{equation}
Therefore, topological 1-soliton solutions to the fKG equation become 
\begin{equation}
	\begin{aligned}
	u_{5,6,7,8}(x,t)= \pm\sqrt{-\frac{p}{q}} \tanh(\pm\sqrt{\nu^2+\frac{1}{2}p}x-\nu\frac{t^{\alpha}}{\alpha})
	\end{aligned}
\end{equation}
where $q\neq 0$.
The graphical illustration of $u_5(x,t)$ for various choices of $p$, $q$ and $\nu$ is given in Fig \ref{fig2a}-\ref{fig2d} for different values of $\alpha$.

\section{Topological 1-Soliton Solutions to the fractional potential Kadomtsev-Petviashvili (fpKP) Equation}
\begin{figure}[H]
 
   \subfigure[$\alpha=0.25$]{
   \includegraphics[scale =0.4] {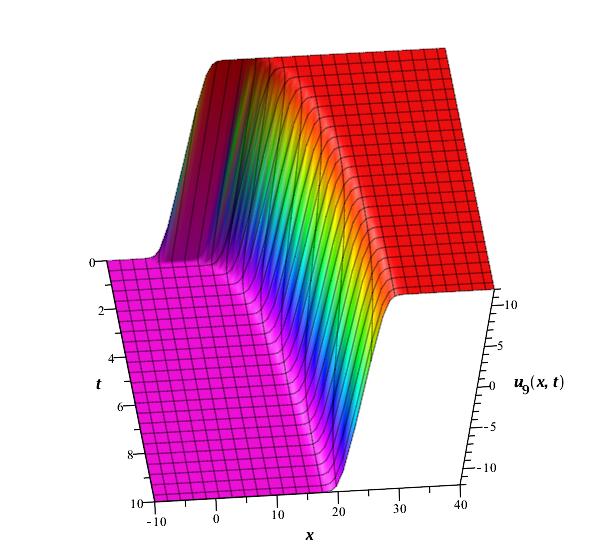}
   \label{fig3a}
 }
  \subfigure[$\alpha=0.50$]{
   \includegraphics[scale =0.4] {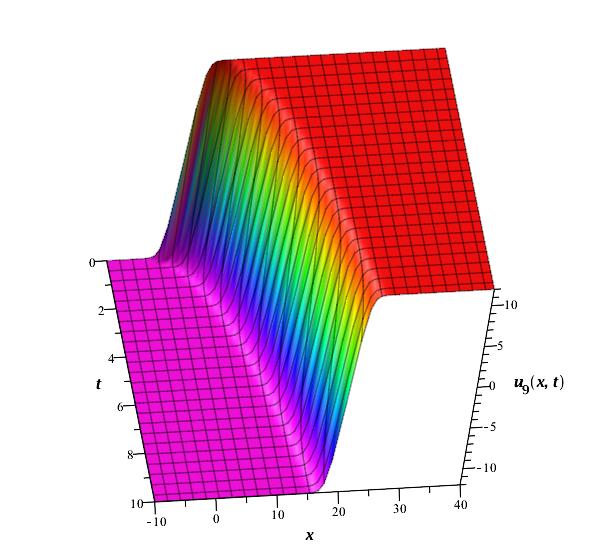}
   \label{fig3b}
 }
 \subfigure[$\alpha=0.75$]{
   \includegraphics[scale =0.4] {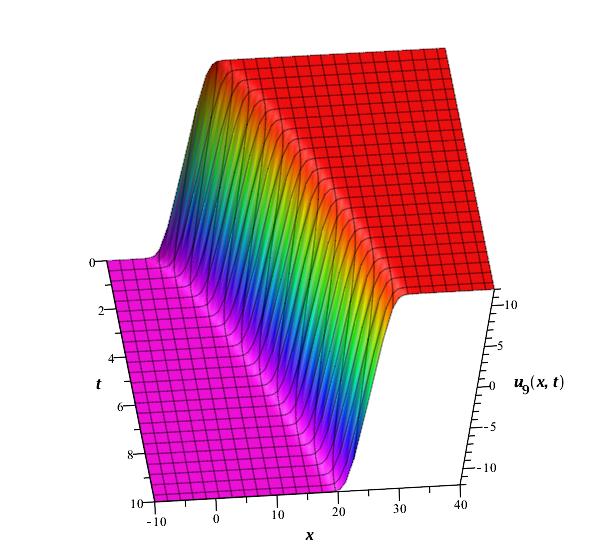}
   \label{fig3c}
 }
 \subfigure[$\alpha=1.00$]{
   \includegraphics[scale =0.4] {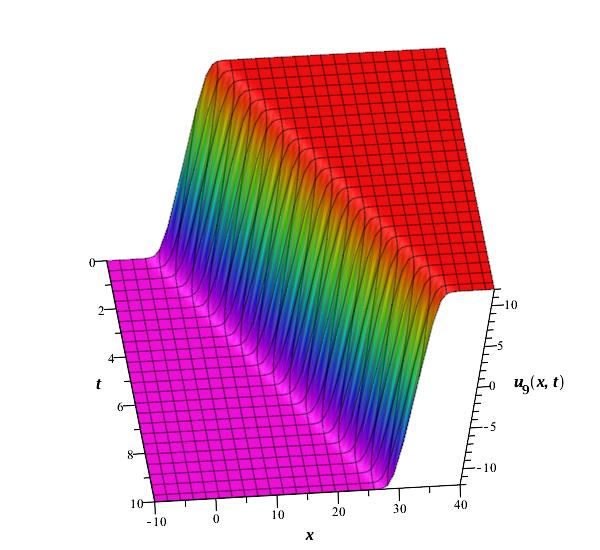}
   \label{fig3d}
 }
 \caption{Projection of the solution $u_9(x,t)$ for $p=q=a=1$, $r=-1$ and $y=0$.}
\end{figure}

The time fractional potential Kadomtsev-Petviashvili (fpKP) equation of the form
\begin{equation}
	D_{t}^{\alpha}u_x+pu_{xx}u_x+qu_{xxxx}+ru_{yy}=0
\end{equation}
is reduced to the ordinary differential equation
\begin{equation}
	-\nu a U''+pa^3U''U'+qa^4U''''+rb^2U''=0 \label{pkp}
\end{equation}
by using the wave transformation (\ref{wt}). Assuming (\ref{pkp}) has a solutions of the form $\tanh^B{\xi}$ and substituting this solution into (\ref{pkp}) gives
\scriptsize{
\begin{equation} 
\begin{aligned}
	 &\left( 6\,A{B}^{4}{a}^{4}q+10\,A{B}^{2}{a}^{4}q-2\,A{B}^{2}{b}^{2}r+2
\,A{B}^{2}a\nu \right)  \tanh^{B}  \xi \\
&+
 \left( A{B}^{4}{a}^{4}q-6\,A{B}^{3}{a}^{4}q+11\,A{B}^{2}{a}^{4}q-6\,A
B{a}^{4}q \right)   \tanh  ^{B-4}\xi \\
&+ \left( -4\,A{B}^{4}{a}^{4}q+12\,A{B}^{3}{a}^{4}q-16\,A{B}^{2}{a}^{4}q
+8\,AB{a}^{4}q+A{B}^{2}{b}^{2}r-A{B}^{2}a\nu-AB{b}^{2}r+ABa\nu
 \right)  \tanh^{B-2}\xi \\
&+ \left( -4\,
A{B}^{4}{a}^{4}q-12\,A{B}^{3}{a}^{4}q-16\,A{B}^{2}{a}^{4}q-8\,AB{a}^{4
}q+A{B}^{2}{b}^{2}r-A{B}^{2}a\nu+AB{b}^{2}r-ABa\nu \right)  
\tanh ^{B+2} \xi \\
&+ \left( A{B}^{4}{a}^{4}q+6\,A
{B}^{3}{a}^{4}q+11\,A{B}^{2}{a}^{4}q+6\,AB{a}^{4}q \right)  
\tanh ^{B+4}\xi \\
&+ \left( {A}^{2}{B}^{3}{a}^{3}
p-{A}^{2}{B}^{2}{a}^{3}p \right)  \tanh^{2\,B-3}  \xi \\
&+ \left( -3\,{A}^{2}{B}^{3}{a}^{3}p+{A}^{2}{B}^{2}{a
}^{3}p \right)   \tanh ^{2\,B-1} \xi  \\
&+ \left( 3\,{A}^{2}{B}^{3}{a}^{3}p+{A}^{2}{B}^{2}{a}^{3}p \right) 
 \tanh^{2\,B+1}  \xi \\
&+ \left( -{A}^{2}{B
}^{3}{a}^{3}p-{A}^{2}{B}^{2}{a}^{3}p \right)  \tanh^{2\,B+3} \xi
 =0
\end{aligned}
\end{equation}}
 \normalsize
Equating the power $2B-3$ to $B-2$ gives $B=1$. Thus, the predicted solution becomes $\tanh{\xi}$. Substituting this solution into (\ref{pkp}) gives
\begin{equation}
\begin{aligned}
	 &\left( -2\,{A}^{2}{a}^{3}p+24\,A{a}^{4}q \right)  \tanh^{5}
 \xi + \left( 4\,{A}^{2}{a}^{3}p-40\,A{a}^
{4}q+2\,A{b}^{2}r-2\,Aa\nu \right)   \tanh ^{3} \xi 
  \\
&+ \left( -2\,{A}^{2}{a}^{3}p+16\,A{a}^{4}q-2\,A{b}^{2}r+2
\,Aa\nu \right) \tanh \xi  =0
\end{aligned}
\end{equation}
Since we seek a nonzero solution, the only coefficients of the powers of $\tanh\xi$ should be zero. Therefore, the algebraic system of equations is obtained

\begin{equation}
	\begin{aligned}
	-2\,{A}^{2}{a}^{3}p+24\,A{a}^{4}q&=0\\
	4\,{A}^{2}{a}^{3}p-40\,A{a}^
{4}q+2\,A{b}^{2}r-2\,Aa\nu&=0\\
-2\,{A}^{2}{a}^{3}p+16\,A{a}^{4}q-2\,A{b}^{2}r+2
\,Aa\nu&=0
	\end{aligned}
\end{equation}
The solutions of this system for $A$, $\nu$, $a$ and $b$ gives
\begin{equation}
	\begin{aligned}
	  A&=\frac{12aq}{p} \\
		\nu&=\frac{4a^4q+b^2r}{a}
	\end{aligned}
\end{equation}
for arbitrarily chosen $a\neq 0$ and $b$ where $p\neq 0$. Thus, the solution to (\ref{pkp}) is constructed as
\begin{equation}
	U(\xi)=\frac{12aq}{p} \tanh{\xi}
	\end{equation}
Returning to the original variables gives the solution to the fpKP as
\begin{equation}
	u_9(x,y,t)=\frac{12aq}{p} \tanh{\left(ax+by-\frac{4a^4q+b^2r}{a}\frac{t^{\alpha}}{\alpha}\right)}
\end{equation} 
where $p\neq 0$ and $a\neq 0$.
The graphical illustration of $u_p(x,t)$ for various choices of $p$, $q$, $r$, $a$ and $b$ is given in Fig \ref{fig3a}-\ref{fig3d} for different values of $\alpha$.

\section{Conclusion}
In the present study, a simple hyperbolic tangent ansatz is used to derive topological 1-soliton solutions to some one and two dimensional fractional nonlinear partial differential equations. The conformable derivative supports the chain rule. Compatible traveling wave transformation reduces the fractional partial differential equations to some ordinary differential equations. The hyperbolic tangent ansatz is a predicted solution and includes some parameters to be determined in an order. The first parameter to be determined is the positive integer power parameter. The determination of this parameter is followed by the other parameters in the solution by solving some algebraic equation systems. The time fractional modified EW, Klein-Gordon and potential Kadomtsev-Petviashvili equations are solved by using the hyperbolic tangent ansatz. Explicit solutions are derived and graphical illustrations are represented in $3D$ plots by assist of computer.

\end{document}